\documentclass{article}

\input{tcilatex}

\begin{document}

\title{Kinetic theory of cluster impingement in the framework of statistical
mechanics of rigid disks.}
\author{M. Tomellini, M. Fanfoni $^{(\ast )}$ \and Dipartimento di Scienze e
Tecnologie \and ChimicheUniversit\`{a} di Roma Tor Vergata \and Via della
Ricerca Scientifica 00133 Roma Italy \and $^{(\ast )}$ Dipartimento di
Fisica Universit\`{a} di Roma Tor Vergata \and Via della Ricerca Scientifica
00133 Roma Italy}
\maketitle

\begin{abstract}
The paper centres on the evaluation of the function $n(\Theta )=N(\Theta
)/N_{0}$, that is the normalized number of islands as a function of coverage 
$\Theta \in \lbrack 0,1]$, given $N_{0}$ initial nucleation centres (dots)
having any degree of spatial correlation. A mean field approach has been
employed: the islands have the same size at any coverage. In particular, as
far as the random distribution of dots is concerned, the problem has been
solved by considering the contribution of binary collisions between islands
only. With regard to correlated dots, we generalize a method previously
applied to the random case only. In passing, we have made use of the
exclusion probability reported in [S. Torquato, B. Lu, J. Rubinstein,
Phys.Rev.A 41, 2059 (1990)], for determining the kinetics of surface
coverage in the case of correlated dots, improving our previous calculation
[M. Tomellini, M. Fanfoni, M. Volpe Phys. Rev.B 62, 11300, (2000)].
\end{abstract}

\section{Introduction}

For describing the evolution of thin film morphologies several physical
processes have to be considered, among others the atom condensation and
evaporation, the adatom surface diffusion, the nucleation and the island
growth. During the last decades significant advances have been done in
modeling, taking advantage of both analytical and numerical approaches, the
inititial stage of film formation, namely the nucleation and\ the growth
processes in the low coverage regime \cite{Rev Zinke Allmang}, \cite{Rev
Brune}. The methods range from the classical mean field rate equations to
the scaling theories which are found to be suitable tools for describing
both experimental and Kinetic Monte Carlo results \cite{Libro Woodruff}. In
order to go beyond the early stage of the growth, it is mandatory to tackle
another physical process, that is the collision among the growing clusters.
In the case of diffusionless clusters this process is actually linked to the
cluster growth only. As a consequence an island which is in general a
collection of connected clusters stems from one or more nucleation events.
As regards the film morphology two mechanisms can be distinguished: the 
\textit{impingement} and the \textit{coalescence}. In the former no
redistribution of matter among clusters occurs after a collisional event and
the clusters retain their individuality. Conversely, in the latter mechanism
redistribution of matter does occur with conservation of both mass and shape.

The process of coalescence has been modeled, for the first time, by Vincent 
\cite{Vincent} who proposed an analytical solution for the island density
kinetics on the basis of the Poisson distribution. Briscoe and Galvin
developed a statistical theory for island coalescence which is in good
agreement with the behaviour of the island density decays obtained by
computer simulations \cite{Briscoe et Galvin}. As far as the impingement
mechanism is concerned, it has been faced in a certain details in ref.\cite%
{Coll Serie ASS}. Specifically, the exact solution for the normalized number
density of islands, as a function of surface coverage $n(\Theta )=\frac{%
N(\Theta )}{N_{0}},$\ where $N(\Theta )$ is the number density of islands at
coverage $\Theta $ and $N_{0}\equiv N(0),$\ has been determined in the form
of a series, the so called \textit{collision series}. This computation holds
in the case of Poissonian simultaneous nucleation and in the entire range of
the surface coverage $(0<\Theta <1\ )$. By means of Monte Carlo simulations
it was also shown that the first three terms well approximate the series 
\cite{Our MC Imping}.

Coalescence and impingement mechanisms of 2D and 3D islands have been
extensively studied, through Monte Carlo simulations, as a function of the
nucleus shape \cite{Our PRB Coal+Imp}. This work clearly demonstrates that,
independently of the collision mechanism governing the film formation, the $%
n(\Theta )$ kinetics is, in fact,the same, in spite of the fact that in the
time domain the two mechanisms give rise to completely different kinetics.
As a consequence, the behaviour of $n(\Theta )$ is expected to be the same
also in the intermediate cases, namely the \textit{partial coalescence }\cite%
{Our PRB Coal+Imp}. The universal behaviour of the $n(\Theta )$ kinetics can
be succesfully exploited for determining, from experimental data, the
nucleation density at saturation. To this end it is sufficient to measure
island densities in the high coverage regime $(\Theta >15-20\%)$, when
islands are sufficiently large and atomic resolution microscopy is not
required. An application of this approach to the growth of both diamond on
Si substrate and quaterthiophene films on silica substrate has been recently
presented in \cite{Polini N(S)},\cite{Borghese}.

Thanks to the universal behaviour of the normalized island density, $%
n(\Theta )=\frac{N(\Theta )}{N_{0}},$ also the non simultaneous nucleation
case can be tackled. To this purpose one resorts to the classical mean field
rate equations for both island and adatom densities. These equations can be
integrated provided the rate coefficient for island collisions is evaluated 
\cite{Venable Phyl Mag}. As a matter of fact this coefficient is a function
of $n(\Theta )$ \cite{Coll Serie ASS}. The universal behavior of this
function, together with the key role it plays in the rate equation approach
to film growth, motivated further analytical studies. Recently, on the
ground of a mean field approximation a semi-analytical approach has been
developed which leads to island density kinetics that is in excellent
agreement with the numerical simulation over the whole range of coverage.
The method leads to \cite{Our Edge-Edge Analitico},

\begin{equation}
n(\Theta )=\left[ \frac{\left( \pi /4\right) ^{1/2}-\Theta ^{1/2}}{%
W_{0}(\Theta )}\right] ^{2},  \label{An Kin 1}
\end{equation}%
where the function $W_{0}(\Theta )=\left( \ln \frac{1}{1-\Theta }\right)
^{1/2}\int_{0}^{\infty }(1-\Theta )^{\xi \left( 1+\xi \right) }d\xi $ has to
be evaluated numerically.

The models discussed so far refer to the case in which nuclei are
distributed at random throughout the whole surface. However, in nucleation
processes ruled by adatom diffusion there exists a zone, around each
nucleus, where the nucleation rate is reduced \cite{Venable Phyl Mag}. In
other words, the nucleation events do not occur at random on the entire
surface and this brings about the establishment of a degree of spatial
correlation among nuclei.

One of the motivation of the present article is therefore the extension of
the aforementioned analytical approach to the more realistic case of
spatially correlated nuclei. Like in the random distribution of nuclei, also
in this case rate equations can be employed to deal with non simultaneous
nucleation. In this instance it is worth noting that rate equations are
suitable for describing Kinetic Monte Carlo results of correlated nucleation 
\cite{OUR APL}. However, for want of theoretical modeling, in \cite{OUR APL}
the $n(\Theta )$ function has been determined numerically through Monte
Carlo simulations (MC) \cite{KMC OUR APL}.

The paper is organised as follows. In the first section we propose a novel
analytical approach based on the statistical mechanics of rigid disks, for
the evaluation of the $n(\Theta )$ kinetics in the random case. The rate
coefficient for island collision is also computed, analytically. In the
second section the method presented in \cite{Our Edge-Edge Analitico} will
be employed to tackle the impingement process in simultaneous nucleation of
spatially correlated nuclei.\bigskip 

\section{Results and discussion}

\subsection{Random distribution of nuclei}

The following analytical approach for computing $n(\Theta )$, is based on a
mean field approximation, in the sense that at any given coverage (time),
all islands have the same shape and appropriate size (disks in the case in
point). This means that we are actually dealing with a sort of coalescence
mechanism, yet, due to the universal behaviour of the $n(\Theta )$ function,
the result of the computation can be applied to the impingement case as
well. Collisions involving more than two islands will be neglected.

Let us denote by $dP$ the probability that an island be involved in a binary
collision. Since we are dealing with binary collision the changing rate of
the island number is just equal to the rate of "dimer" formation; this rate
is equal to half the rate at which \textit{an island} undergoes a collision
event. Therefore the relation holds

\begin{equation}
\frac{dN}{N}=-\frac{1}{2}dP,  \label{binaria 1}
\end{equation}%
where $N$ stands for the number density of islands. A simple closure of eqn.%
\ref{binaria 1} is achieved by setting $dP=2\pi N(2R)d(2R),$ $R$ being the
radius of the islands, that is by considering the radial distribution
function of the disks to be equal to one. For $\Theta =N\pi R^{2},$ to a
first approximation one receives $\frac{dN}{N}\simeq -2d\Theta $ or \cite%
{Hutchinson} 
\begin{equation}
n(\Theta )=e^{-2\Theta }.  \label{An  G=1}
\end{equation}

As will soon be clear this is a poor description of $n(\Theta )$ over the
entire range of coverage, it describes the kinetics in the low coverage
regime only. In fact, in the limit $\Theta \rightarrow 0$ the kinetics
becomes $n(\Theta )\sim 1-2\Theta $ as already derived in\ \cite{Our PRB
Coal+Imp}.

The reason of this inaccuracy relies in the fact that the radial
distribution function of a system of impenetrable disks is not equal to one
except in the limit of large particle separation. A more suitable choice of
the probability for binary collision is indispensable in order to model the
island density kinetics over the whole range of coverage. To this end the
results achieved on the thermodynamic system of rigid spheres \cite{Lebowitz}%
, \cite{Hansen}, \cite{Torquato 1} can be properly exploited. In particular
we will make use of the results and the notation of ref.\cite{Torquato 1}.

$E_{v}(r)$ is the probability of finding a region of area $\pi r^{2}$
centered at some arbitrary point empty of island centers; $E_{p}(r)$ is the
probability that given an island (its center) at some arbitrary point, the
region of area $\pi r^{2}$ encompassing the central island is empty of
island centers. The respective density probability functions are attained by
the derivatives of the exclusion probabilities $E_{v}(r)$ and $E_{p}(r)$: $%
H_{v}(r)=-\frac{\partial E_{v}(r)}{\partial r}$ and $H_{p}(r)=-\frac{%
\partial E_{p}(r)}{\partial r}$. The expression of the exclusion probability 
$E_{v}(r),$ for a system of particles not necessarely correlated through
hard core potentials, has been derived in refs. \cite{Torquato 1}, \cite%
{Giurassico} .

On the ground of the definition of the exclusion probability one gets $%
dP=H_{p}(\sigma )d\sigma ,$ where $\sigma =2R$ is the average value of the
island diameter. The rate equation for binary collision then becomes

\begin{equation}
\frac{dN}{N}=-\frac{1}{2}H_{p}(\sigma )d\sigma .  \label{Binaria 2}
\end{equation}

In the thermodynamic limit, "voids" and "islands" exclusion probabilities
are linked by the relationship \cite{Torquato 1}

\begin{equation}
E_{p}(r)=\frac{E_{v}(r)}{E_{v}(\sigma )},  \label{Torq 1}
\end{equation}%
which holds for $r\geq \sigma .$ Eqn.\ref{Torq 1} shows that for a system of
hard disks $E_{p}(\sigma )=1.$ Furthermore in the same limit \cite{Torq a1a0}

\begin{equation}
H_{p}(r)=\frac{8\Theta }{\sigma }\left( a_{0}\frac{r}{\sigma }-a_{1}\right)
E_{p}(r/\sigma ).  \label{binaria3}
\end{equation}%
where $a_{0}=\frac{1+b_{0}\Theta }{\left( 1-\Theta \right) ^{2}}$, $a_{1}=-%
\frac{b_{1}\Theta }{\left( 1-\Theta \right) ^{2}}$ and $E_{p}(x)=e^{-\Theta %
\left[ 4a_{0}(x^{2}-1)+8a_{1}(x-1)\right] }.$

By using eqn.\ref{binaria3} in the kinetic equation eqn.\ref{Binaria 2}, we
eventually get

\begin{equation}
\frac{dN}{N}=-\frac{2\left( 1-a\Theta \right) }{(1-\Theta )^{2}}d\Theta =-%
\frac{d\Theta }{\tau _{c}(\Theta )},  \label{binaria 4}
\end{equation}%
where $d\Theta =\frac{1}{4}\pi Nd(\sigma ^{2})$ is the increment of surface
fraction covered by islands, $a=b_{1}-b_{0}$ and $\tau _{c}(\Theta )$ the
characteristic collision-time function. The definition of the time constant
as given by eqn.\ref{binaria 4} is in fact required in order to treat the
non simultaneous nucleation by menas of rate equations \cite{Current Opinion}%
,\cite{OUR APL}. The integration of eqn.\ref{binaria 4} can be performed
analitically and gives,

\begin{equation}
n(\Theta )=\left( 1-\Theta \right) ^{2a}e^{-\frac{2(1-a)\Theta }{1-\Theta }}.
\label{Islan kin1}
\end{equation}

The validity of our approach has been tested by using Monte Carlo (MC)
simulations of island density decays in film growth ruled by the impingement
mechanism \cite{Our PRB Coal+Imp}. In fig.1 eqn.\ref{Islan kin1} (for $%
a=0.564-0.128=0.436$ \cite{Torq a1a0}) has been compared to the MC
simulation. Remarkably, the agreement between the simulation and the
analytical result is excellent. In any case also the kinetics for $%
a=0.5-0=0.5$ \cite{Hefland}\ is in very good agreement with the simulation
(not shown).

Another analytical but less precise approach, is based on the exclusion
probability already derived by us in ref.\cite{OUR CORR delta}. By only
retaining the Heaviside contribution in the radial distribution function of
hard disks and decoupling the multiple integral which gives the argument of
the exponential of the $E_{v}(r)$ function \cite{OUR CORR delta}, we found

\begin{equation}
E_{v}(r)=e^{-\pi Nr^{2}\left[ 1+\frac{1}{2}\pi Nr^{2}\chi _{\lbrack 0,\sigma
]}(r)+\frac{1}{2}\pi N\sigma ^{2}\chi _{(\sigma ,\infty )}(r)\right] },
\label{pr vuot our}
\end{equation}%
where $\chi _{A}(x)$ is equal to one if $x\in A$ and is equal to zero if $%
x\notin A$. It is worth noting that this expression has been successfully
employed for modeling the kinetics of the surface coverage in case of
spatially correlated nucleation [see the end of the article and \cite{OUR
CORR delta}]. From eqn.\ref{pr vuot our} on account of eqn.\ref{Torq 1} we
get 
\begin{equation}
H_{p}(\sigma )=-\frac{\left. \partial _{r}E_{v}(r)\right\vert _{\sigma }}{%
E_{v}(\sigma )}=-\left. \frac{\partial \ln E_{v}(r)}{\partial r}\right\vert
_{\sigma }=2\pi N\sigma \left( 1+\frac{3}{4}N\pi \sigma ^{2}\right) ,
\label{Gp OUR}
\end{equation}%
that, once inserted in eqn.\ref{Binaria 2}, yields the kinetics

\begin{equation}
n(\Theta )=e^{-(2\Theta +3\Theta ^{2})}.  \label{NsuN0 GM}
\end{equation}%
As shown in fig.1 this kinetics is in a pretty good agreement with the MC
simulation over the entire range of coverage \cite{Nota 0}. In the same
figure it is also displayed the behavior of the solution eqn.\ref{An G=1}.
As appears, the modeling based on this last approximation is inadequate for
describing the kinetics in the whole range of surface coverage; in fact it
is in agreement with the MC output only at very beginning of the kinetics.

\subsection{Spatially correlated nuclei}

The kinetics of island density in the case of Dirac delta (or heterogeneous)
correlated nucleation can be determined by the same method as that employed
in ref.\cite{Our Edge-Edge Analitico}. Although the exclusion probability
computed in ref.\cite{Giurassico} allows, in principle, to treat any kind of
correlation, here it is introduced in such a way that nucleation in a
circular region of radius $R_{hc}$ around each nucleus is prevented. It goes
without saying that during the growth the radius of the cluster can exceed $%
R_{hc}.$ Let us introduce the extended surface, $\Theta _{e}=\pi N_{0}R^{2},$
where $R$ stands for the nucleus radius and the quantity $\Theta ^{\ast
}=\pi N_{0}R_{hc}^{2}$ as a measure of the correlation degree of the system.
The average value of the edge-to-edge distance among islands is given in
terms of the exclusion probability as follows

\begin{eqnarray}
\bar{z}(R;\Theta ^{\ast }) &=&\frac{2\dint\limits_{0}^{\infty
}E_{v}(z+R;\Theta ^{\ast })dz}{E_{v}(R;\Theta ^{\ast })}  \nonumber \\
&=&\frac{2R\dint\limits_{0}^{\infty }E_{v}\left[ R(1+\xi );\Theta ^{\ast }%
\right] d\xi }{1-\Theta (R;\Theta ^{\ast })},  \label{Edge-edge}
\end{eqnarray}%
where in the last equation the identity $E_{v}(R;\Theta ^{\ast })=1-\Theta
(R;\Theta ^{\ast })$ has been exploited and $\xi =z/R;$ clearly $R\neq 0$.
Therefore the eqn.\ref{Edge-edge} reads

\begin{equation}
\bar{z}(\Theta ;\Theta ^{\ast })=\frac{2F(\Theta ;\Theta ^{\ast })}{\sqrt{%
\pi N_{0}}},  \label{dge-edge 2}
\end{equation}%
with 
\begin{equation}
W_{1}(\Theta ;\Theta ^{\ast })=\frac{\Theta _{e}^{1/2}}{1-\Theta }%
\dint\limits_{0}^{\infty }E_{v}\left[ R(1+\xi );\Theta ^{\ast }\right] d\xi .
\label{edge-edge 3}
\end{equation}

The average diameter of the island when the fraction of covered surface is $%
\Theta $ and $N$ is the island density reads

\begin{equation}
D(\Theta ,\Theta ^{\ast })=2\left( \frac{\Theta }{\pi N(\Theta ,\Theta
^{\ast })}\right) ^{1/2},  \label{Island Diameter}
\end{equation}%
while the distance between the centers of the islands can be written as

\begin{equation}
\bar{d}(\Theta ,\Theta ^{\ast })=\frac{C}{\sqrt{\pi N(\Theta ,\Theta ^{\ast
})}}.  \label{dist madia isole}
\end{equation}%
where $C$ is a constant to be determined. Apparently,

\begin{equation}
\bar{d}(\Theta ,\Theta ^{\ast })=D(\Theta ,\Theta ^{\ast })+\bar{z}(\Theta
,\Theta ^{\ast })  \label{consvdistanza}
\end{equation}%
and, since for $\Theta =\Theta ^{\ast }/4$, $N=N_{0},$ from eqn.\ref%
{consvdistanza} the $C$ constant is easily determined as $C=2W_{1}(\Theta
^{\ast }/4;\Theta ^{\ast })+\sqrt{\Theta ^{\ast }}$. Using eqns.\ref%
{dge-edge 2}, \ref{Island Diameter}, \ref{dist madia isole} and \ref%
{consvdistanza} we end up with

\begin{equation}
n(\Theta ,\Theta ^{\ast })=\left[ \frac{C-\Theta ^{1/2}}{W_{1}(\Theta
,\Theta ^{\ast })}\right] ^{2}.  \label{N/N0 Correl gen}
\end{equation}%
which holds for $\Theta \geq \Theta ^{\ast }/4.$ As anticipated, we perform
the computation of this kinetics by resorting to the exclusion probability
of ref.\cite{Hefland} which holds for the hard core correlation \cite{Nota1}%
. As far as this expression is concerned, it is given by

\begin{equation}
E_{v}(x;y)=\left[ 1-yx^{2}\right] \chi _{\lbrack 0,1/2)}(x)+(1-\tfrac{y}{4}%
)e^{-\frac{y/4}{(1-y/4)^{2}}(4x^{2}-xy+\frac{y}{2}-1)}\chi _{\lbrack
1/2,\infty )}(x),  \label{Vuoto Hefland1}
\end{equation}%
where $x=(\Theta _{e}/\Theta ^{\ast })^{1/2}(1+\xi )$ and $y=\Theta ^{\ast
}. $ Inserting eqn.\ref{Vuoto Hefland1} in eqn.\ref{edge-edge 3}, the
numerical integration allows one to determine the decay of the islands as
shown in the fig.2 together with the MC simulation. The agreement is
satisfactory.

Before concluding this paper, we take the occasion to apply the exclusion
probability as derived in ref.\cite{Torq a1a0} to the determination of the
kinetics of surface coverage in the case of Dirac delta spatially correlated
nucleation. In our previous article \cite{OUR CORR delta} we evaluated the
kinetics at hands by using eqn.\ref{pr vuot our} for three values of $\Theta
^{\ast }$, namely 0.2, 0.7 and 1.5. Although those results were quite
satisfactory, they can be improved by using the following exclusion
probability \cite{Torq a1a0}

\begin{equation}
E_{v}(\Theta _{e};\Theta ^{\ast })=(1-\Theta _{e})\chi _{\left[ 0\right.
,\left. \frac{\Theta ^{\ast }}{4}\right) }+\left( 1-\frac{\Theta ^{\ast }}{4}%
\right) e^{-\frac{\Theta ^{\ast }}{4}\gamma (\Theta ^{\ast })\left[ 4\frac{%
\Theta _{e}}{\Theta ^{\ast }}-\sqrt{\Theta ^{\ast }\Theta _{e}}+\frac{\Theta
^{\ast }}{2}-1)\right] }\chi _{\left[ \frac{\Theta ^{\ast }}{4}\right.
,\left. \infty \right) }  \label{Ev Torq}
\end{equation}

where $\gamma (\Theta ^{\ast })=\frac{1+0.128\Theta ^{\ast }/4}{(1-\Theta
^{\ast }/4)^{2}}.$ The results are displayed in fig.3 together with the MC
kinetics which have been described in ref.\cite{OUR CORR delta}.

\QTP{Body Math}
\newpage

\section{Figure captions}

Fig.1) Kinetics of island impingement as a function of surface coverage in
case of simultaneous nucleation of randomly distributed nuclei. The Monte
Carlo result and the analytical solution (eqn.8) are shown as dots, and full
line, respectively. The solution eqn.11 is reported as dashed line. The
behaviour of eqn.3 is displayed as open circles.

Fig.2) Kinetics of island impingement as a function of surface coverage in
case of simultaneous nucleation of spatially correlated nuclei. Symbols are
the Monte Carlo results for a correlated system according to the hard-core
model for $\Theta ^{\ast }=0.8.$ The semi analytical result is shown as full
line (eqn.18)

Fig.3) Kinetics of surface coverage fraction as a function of the extended
surface, $\Theta _{e}$, for simultaneous nucleation of correlated nuclei
according to the hard-core model. The Monte Carlo simulations and the
analytical results are shown as symbols and full lines, respectively. Curves
a, b and c refers to $\Theta ^{\ast }=0.2,$ $\Theta ^{\ast }=0.7$, $\Theta
^{\ast }=1.5$, respectively.


\newpage

\begin{thebibliography}{99}
\bibitem{Rev Zinke Allmang} M. Zinke-Allmang, L.C. Feldman, M.C. Grabow,
Surf. Sci. Rep. 16 (1992) 337

\bibitem{Rev Brune} H. Brune, Surf. Sci. Rep. 31 (1998) 121

\bibitem{Libro Woodruff} Growth and properties of Ultra Thin Epitaxial Layer
in Chemical Physics of Solid Surfaces, vol 8 D.a. King, D.P. Woodruff (Edrs)
Elsevier 1997

\bibitem{Vincent} R. Vincent, Proc. R. Soc. London Ser.A 321 (1971) 53

\bibitem{Briscoe et Galvin} B.J. Briscoe, K.P. Galvin, Phys. Rev. A 43
(1991) 1906

\bibitem{Coll Serie ASS} M. Fanfoni, M. Tomellini, Appl. Surf. Sci. 136
(1998) 338

\bibitem{Our MC Imping} M. Volpe, M. Fanfoni, M. Tomellini, V. Sessa, Surf.
Sci. Lett. 440 (1999) L820

\bibitem{Our PRB  Coal+Imp} M. Fanfoni, M. Tomellini, M. Volpe, Phys. Rev. B
64 (2001) 075409

\bibitem{Polini N(S)} M. Fanfoni, R. Polini, V. Sessa, M. Tomellini, M.
Volpe, Appl. Surf. Sci. 152 (1999) 126

\bibitem{Borghese} M. Campione, A. Borghese, M. Moret, A. Sassella, J.
Mater. Chem. 13 (2003) 1669

\bibitem{Venable Phyl Mag} J. A. Venable, Phyl. Mag. 17 (1973) 697

\bibitem{Our Edge-Edge Analitico} M. Tomellini, M. Fanfoni,Surf. Sci. 450
(2000) L267

\bibitem{OUR APL} M. Fanfoni, M. Tomellini, M. Volpe, Appl. Phys. Lett. 78
(2001) 3424

\bibitem{KMC OUR APL} H. Brune, G.S. Bales, J. Jacobsen, C. Boragno, K.
Kern, Phys. Rev. B 60 (1999) 5991

\bibitem{Hutchinson} M.J. Stowell, T.E. Hutchinson, Thin Solid Film 8 (1971)
41

\bibitem{Our N. Cimento} M. Fanfoni, M. Tomellini, Il Nuovo Cimento 20
(1998) 1171

\bibitem{Lebowitz} H. Reiss, H.L. frisch, J.L. Lebowitz, J. Chem. Phys. 31
(1939) 369

\bibitem{Hansen} J.P. Hansen, I.R. Mc Donald, Theory of Simple Liquids, 2nd
ed. (Academic, London, 1986)

\bibitem{Torquato 1} S. Torquato, B. Lu, J. Rubinstein, Phys. Rev. A 41
(1990) 2059

\bibitem{Giurassico} M. Fanfoni, M. Tomellini, Eur. Phys. J. B 34 (2003) 331

\bibitem{Torq a1a0} S. Torquato, Phys. Rev. E51 (1995) 3170

\bibitem{Current Opinion} M. Tomellini, M. Fanfoni, Current Opinion in Solid
State and Materials Science, 5 (2001) 91; M. Fanfoni, M. Tomellini, J.
Phys.:Condens. Matter 17 (2005) R571

\bibitem{Hefland} E. Hefland, H.L. Frisch, J.L. Lebowitz, J. Chem. Phys. 34
(1961) 1037

\bibitem{OUR CORR  delta} M. Tomellini, M. Fanfoni, M. Volpe, Phys. Rev. B
62 (2000) 11300

\bibitem{Nota 0} Clearly, the Monte Carlo simulation which gives the
normalized number of islands is reproducible through an appropriate fit. It
is worth noting that $n(\Theta )=e^{-(2.2\Theta +5.22\Theta ^{3})}$
perfectly fit to the MC simulation.

\bibitem{Nota1} We made use of that expression just for simplicity, yet the
exclusion probability derived in \cite{Torq a1a0} can be employed as well.
\end{thebibliography}
\end{document}